\documentclass[manuscript]{aastex}          
\usepackage{spr-astr-addons}    

\begin{document}
%
\title{Interactions between brown-dwarf binaries and Sun-like stars}

\shorttitle{Brown-dwarf binaries and Sun-like stars}
\shortauthors{M. Kaplan et al.}

\author{M. Kaplan} 
\affil{Department of Space Sciences and Technologies, Akdeniz University, 07058 Antalya,Turkey}
\email{muratkaplan@akdeniz.edu.tr} 

\author{D. Stamatellos}
\affil{School of Physics and Astronomy, Cardiff University, Queens Buildings, The Parade, Cardiff, CF24 3AA, Wales, UK}

\author{A. P. Whitworth}
\affil{School of Physics and Astronomy, Cardiff University, Queens Buildings, The Parade, Cardiff, CF24 3AA, Wales, UK}


\begin{abstract}
Several mechanisms have been proposed for the formation of brown dwarfs, but there is as yet no consensus as to which -- if any -- are operative in nature. Any theory of brown dwarf formation must explain the observed statistics of brown dwarfs. These statistics are limited by selection effects, but they are becoming increasingly discriminating. In particular, it appears (a) that brown dwarfs that are secondaries to Sun-like stars tend to be on wide orbits, $a\ga 100\,{\rm AU}$ (the Brown Dwarf Desert), and (b) that these brown dwarfs have a significantly higher chance of being in a close ($a\la 10\,{\rm AU}$) binary system with another brown dwarf than do brown dwarfs in the field. This then raises the issue of whether these brown dwarfs have formed {\it in situ}, i.e. by fragmentation of a circumstellar disc; or have formed elsewhere and subsequently been captured. We present numerical simulations of the purely gravitational interaction between a close brown-dwarf binary and a Sun-like star. These simulations demonstrate that such interactions have a negligible chance ($<0.001$) of leading to the close brown-dwarf binary being captured by the Sun-like star. Making the interactions dissipative by invoking the hydrodynamic effects of attendant discs might alter this conclusion. However, in order to explain the above statistics, this dissipation would have to favour the capture of brown-dwarf binaries over single brown-dwarfs, and we present arguments why this is unlikely. The simplest inference is that most brown-dwarf binaries -- and therefore possibly also most single brown dwarfs -- form by fragmentation of circumstellar discs around Sun-like protostars, with some of them subsequently being  ejected into the field.
\end{abstract}

\keywords{stellar dynamics : numerical methods : star formation}

\section{Introduction} \label{S:INTRO}

The existence of brown dwarfs was first suggested  by Kumar (1961), on the basis of theoretical calculations of the structure of newly-formed stars, and the realisation that stars of sufficiently low mass, as they condense out of the interstellar medium, become dense enough to be supported by electron degeneracy pressure before they become hot enough to burn hydrogen. However, it was not until 1995 (Rebolo et al. 1995, Nakajima et al. 1995) that brown dwarfs were actually observed. Since then brown dwarfs have been observed in increasing numbers, and in increasing detail, so that we now have some idea of their statistical properties -- albeit subject to strong selection effects due to the fact that brown dwarfs are intrinsically faint, and become ever fainter with age (Luhman et al. 2007). It is therefore appropriate to evaluate how the proposed mechanisms for brown dwarf formation measure up against the observational constraints.

In discussing the formation of brown dwarfs, it is inappropriate to distinguish them from hyd\-ro\-gen--burning stars, since there is no reason to suppose that the processes that determine the mass of a star condensing out of interstellar matter change radically at the hydrogen-burning limit ($\sim 0.08\,{\rm M}_\odot$). Therefore brown dwarfs are likely to form a continuum with low-mass hydrogen--burning stars. This does not mean that brown dwarfs form in exactly the same way as all other stars, but simply that any systematic changes in going from massive star formation to brown dwarf formation are part of a continuous trend which also takes in the formation of intermediate- and low-mass hydrogen-burning stars\footnote{We note that this is a much less extreme view than that espoused by Thies \& Kroupa (2007, 2008), who argue that there is an abrupt and discontinuous change in the IMF at or just above the brown-dwarf limit.}. Indeed, we propose that, as one goes to lower masses, an increasing fraction of stars is formed by fragmentation of protostellar accretion discs.

\subsection{Formation mechanisms}

The three main mechanisms for brown dwarf formation currently being advocated are the following. Other mechanisms have been proposed (Whitworth et al. 2007), for example the  photo-erosion of a pre-existing core (Whitworth \& Zinnecker, 2005), but are unlikely to be major contributors.

\subsubsection{Turbulent generation of very low-mass prestellar cores.}

In this mechanism, brown dwarfs can form in isolation from exceptionally small and dense prestellar cores. It is argued that interstellar turbulence occasionally delivers compressive flows of sufficient strength to form such cores (e.g. Padoan \& Nordlund 1999; Hennebelle \& Chabrier 2008, 2009). We note that a prestellar core of mass $M$, with isothermal sound speed $a$, only collapses if the external pressure exceeds
\begin{eqnarray}
P_{_{\rm MIN}}&\simeq&\frac{30\,a^8}{G^3\,M^2}\,,
\end{eqnarray}
For a core at the opacity limit, $M\sim 0.003\,{\rm M}_\odot$, with $a\sim 0.2\,{\rm km}\,{\rm s}^{-1}$ (corresponding to molecular gas at $T\sim 10\,{\rm K}$), this yields\footnote{Note that, instead of dividing out the Boltzmann constant to give the pressure in terms of a number-density times a temperature (as is routinely done), we have divided out the mean mass per hydrogen molecule -- which for molecular gas with a hydrogen mass fraction $X=0.70$ is $\bar{m}_{_{{\rm H}_2}}=2m_{\rm p}/X\simeq 4.8\times 10^{-24}\,{\rm g}$ -- so as to give the pressure in terms of the number-density of H$_2$ times a squared velocity. This highlights the very large H$_2$ number densities and velocities required to deliver a ram pressure of this order.}
\begin{eqnarray}
\frac{P_{_{\rm MIN}}}{\bar{m}_{_{{\rm H}_2}}}&\equiv&0.13\times 10^{10}\,\left({\rm H}_2\,{\rm cm}^{-3}\right)\,\left({\rm km}\,{\rm s}^{-1}\right)^2\,;
\end{eqnarray}
 for a core at the brown dwarf limit, it gives
\begin{eqnarray}
\frac{P_{_{\rm MIN}}}{\bar{m}_{_{{\rm H}_2}}}&\equiv&0.20\times 10^{7}\,\left({\rm H}_2\,{\rm cm}^{-3}\right)\,\left({\rm km}\,{\rm s}^{-1}\right)^2\,.
\end{eqnarray}
There is presently no observational evidence for the dense, supersonic, coherently converging flows required to deliver these high ram pressures. Moreover, this mechanism has not been simulated, and therefore there are no predictions for the statistics of discs, outflows, and multiplicity to enable a comparison with observation.

\subsubsection{Collapse and fragmentation of larger prestellar cores.}

In this mechanism, brown dwarf stars form collectively along with hydrogen-burning stars, during the collapse and fragmentation of more massive prestellar cores (i.e. cores which spawn more than one protostar). Brown dwarfs are then delivered to the field by dynamical ejection (Reipurth \& Clarke, 2001; Bate, Bonnell \& Bromm, 2003; Goodwin, Whitworth \& Ward-Thompson, 2004a,b; Bate 2009; Offner et al. 2009, 2010). Only the simulations by Offner et al. (2009) take proper account of radiative feedback from newly-formed protostars, and, although they form brown dwarfs, (a) these stars are still accreting at the end of the simulation, so their final masses are uncertain, and (b) the brown dwarfs formed are too few (i.e. 2) to allow statistical (as distinct from phenomenological) inferences to be made. It is presently unclear whether this mechanism can deliver the relative number of brown dwarfs found in nature (i.e. the stellar IMF), and it offers no explanation for the rather particular binary statistics of brown dwarfs.

\subsubsection{Fragmentation of protostellar accretion discs.}

In this mechanism, a primary star forms attended by an accretion disc. Material infalls onto this disc and spirals into the primary protostar at a rate determined by how quickly it can redistribute angular momentum. However, if the disc becomes sufficiently massive, extended and cool, it fragments to produce secondary stars. These secondary stars are usually of low mass, i.e. very low-mass H-burning stars (VLMSs), brown-dwarf stars and planetary-mass stars (planemos). This mechanism has the merit that it has been simulated numerically in detail, and reproduces the critical statistical properties of brown dwarfs (Whitworth \& Stamatellos, 2006; Stamatellos, Hubber \& Whitworth, 2007; Stamatellos \& Whitworth, 2009). The VLMSs tend to end up closer to the primary (due to a combination of scattering and secular migration), most of the planemos and brown dwarfs are scattered out of the system. Most of the secondaries are attended by small accretion discs -- large enough to fuel signatures of ongoing accretion, but too small to increase the masses of the secondaries significantly. Many of the secondaries are in close binary systems. Many of these close binary systems remain on wide orbits round the primary, but some are ejected and survive the ejection process. Offner et al. (2010) have shown that when radiative feedback is included in their simulations, disc fragmentation is suppressed, due to the large accretion luminosity from the primary protostar. However, Stamatellos, Hubber \& Whitworth (2011) show that when the episodic nature of accretion is taken into account, there are low-luminosity periods when disc fragmentation is likely to occur.

\subsection{Observational constraints on the properties of brown dwarfs}

\subsubsection{The collective properties of brown dwarfs}

There is a growing body of evidence suggesting that brown dwarfs form in similar ways to low-mass hydrogen -- burning stars. The Initial Mass Functions for low-mass hydrogen-burning stars and for brown dwarfs appear to merge continuously at the hydrogen-burning limit (e.g. Moraux et al. 2007; Lodieu et al. 2009). Both young brown dwarfs and young low-mass hydrogen-burning stars are attended by discs (e.g. Luhman et al. 2008), show signatures of accretion, at rates which follow the approximate relationship $\dot{M}\propto M^2$ (e.g. Natta et al. 2004; Mohanty, Jayawardhana \& Basri, 2005; Comer{\'o}n, Testi \& Natta 2010), and drive jets and outflows (e.g. Whelan et al. 2007). In general, young brown dwarf stars appear to be clustered in the same way as hydrogen-burning stars (e.g. Luhman, 2006).

\subsubsection{The binary statistics of brown dwarfs}

As well as reproducing the collective properties of individual brown dwarfs, a theory of brown dwarf formation must also explain their binary statistics. There are three clear trends emerging here. (i) Brown dwarfs that are secondaries to Sun-like stars tend to be on quite wide orbits (separations $S\ga 100\,{\rm AU}$); this is the Brown Dwarf Desert (McCarthy \& Zuckerman, 2004; Sahlmann et al., 2011). (ii) Brown dwarfs that are in binary systems with other brown dwarfs ($\sim 10-20\%$ of them) tend to be on close orbits (typically $S\la 20\,{\rm AU}$ and $\bar{S}\sim 5\,{\rm AU}$) and to have components of comparable mass ($q\sim 1$) (e.g. Bouy et al., 2003; Burgasser et al., 2003; Close et al., 2003; Ahmic et al., 2007; Reid et al., 2008; Gelino \& Burgasser, 2010; Todorov, Luhman \& McLeod, 2010). (iii) A brown dwarf that is a secondary in a wide orbit around a Sun-like star is estimated to be at least twice as likely to be in a close binary with another brown dwarf (and -- approximately -- five times as likely to be in a triple, and thirteen times as likely to be in a quadruple, with other brown dwarfs) as a brown dwarf in the field (e.g. Burgasser, Kirkpatrick \& Lowrance, 2005; Faherty et al., 2010; Faherty et al., 2011). As long as these statistical inferences hold up in the face of improved observational data and analysis, and larger sample sizes, they constitute very discriminating constraints, which theories of the formation of brown dwarfs should seek to address.

The fact that the observations are biased towards younger brown dwarfs (because younger brown dwarfs are brighter), is not a critical issue in this context, since it weights the statistics towards those properties of brown dwarfs that reflect how they are formed, as distinct from those that are produced by subsequent longer-term environmental and evolutionary effects.

\subsection{The origin of brown-dwarf binaries}

If brown dwarfs form from very low-mass prestellar cores, close brown-dwarf binaries with appropriate masses and separations may form by secondary fragmentation when molecular hydrogen dissociates, as argued by Whitworth \& Stamatellos (2006). However, this mechanism has never been demonstrated to work in numerical simulations (e.g. Bate, 1998). Moreover, if it does work in nature, it only delivers isolated brown-dwarf binaries, and there remains a problem with explaining the brown-dwarf binaries that are found in wide orbits round Sun-like stars. The results we report here suggest that, at least in the absence of hydrodynamic dissipation, the capture of close brown-dwarf binaries into wide orbits round Sun-like stars is very unlikely.

If a brown dwarf forms along with other stars from a single core, the subsequent evolution is highly chaotic, and a range of outcomes is possible. The resulting stars tend to define themselves (in terms of their masses) during the highly dissipative phase when the gas switches from isothermality to adiabaticity at densities $\rho\sim 10^{-13}\,{\rm g}\,{\rm cm}^{-3}$. These high densities may not arise exclusively near the centre of mass of the core, but also in filaments channeling material into the centre (e.g. Walch et al. 2010; Girichidis et al. 2010), and it is unclear where conditions will be most favourable for forming brown dwarfs. However, whether they form near the centre (where the densities are higher) or in a filament away from the centre (where the temperatures are lower), they quickly end up in the centre. This paper demonstrates that, if the violent interactions with other stars that then occur are purely gravitational, they are unlikely to lead to the formation and survival of brown-dwarf binaries, and even less likely to deliver such binaries into wide orbits around more massive stars. 

If brown dwarfs form by disc fragmentation, their statistical properties appear to meet all the critical observational constraints (Stamatellos \& Whitworth 2009). 

\begin{itemize}

\item{Discs provide an ideal place to form objects close to the opacity limit. The requirement that the cooling timescale in a fragment be less than the dynamical timescale (Gammie, 2001), which is, in effect, the opacity limit, is more easily met in a disc, since material is parked in the disc and can cool quasistatically until it becomes Toomre unstable (Toomre 1964). By contrast, in a violently compressive turbulent flow, the dynamical timescale, and hence the time available for cooling, are by definition shorter, and so an adiabatic bounce is more likely.}

\item{Disc fragmentation naturally produces protostars with low masses and a mass-function $d{\cal N}/dM\propto M^{-0.6}$.}

\item{The distribution of angular momentum in a disc ensures that the reservoir of material that a particular protostar can accrete is limited, so there is much less need to cut off accretion by dynamical ejection. Ejections do occur, but there is no need for them to occur rapidly.}

\item{Fragmentation only occurs in the outer reaches of a disc ($R\ga 100\,{\rm AU}$). The VLMSs that form by fragmentation tend to be scattered or migrate into closer orbits, but the brown dwarfs and planemos tend to be scattered outwards and/or ejected.}

\item{Protostars formed by disc fragmentation are attended by their own small accretion discs, with masses and radii comparable with those inferred from observation.}

\item{Protostars condensing out of discs have a natural tendency to pair up and form close binaries, and these close  binaries tend to be on wide orbits around the primary (Sun-like) star.}

\item{Many of the brown dwarfs and planemos are dynamically ejected. However, since these ejections involve rather mild interactions with other low-mass fragments, accretion discs and close low-mass companions can survive the ejection process.}

\end{itemize}

It has been argued (a) that when radiative feedback from the central primary protostar is included, disc fragmentation is suppressed (Offner et al. 2009, 2010), and (b) that fragmenting discs are not seen (Maury et al. 2010). It is therefore appropriate to comment briefly on these issues. (a) Stamatellos, Hubber \& Whitworth (2011) show that if accretion onto the central primary protostar is episodic -- and there is considerable evidence to suggest that this is the case -- then there are periods during which the accretion rate, and hence the luminosity, are low, and these periods are sufficiently long to allow the disc to become very cool and fragment. (b) The timescale for dynamical fragmentation is very short for brown dwarfs, $t_{_{\rm DYN}}\sim 6\,{\rm kyr}\,\left(M/0.01{\rm M}_\odot\right)$. Therefore the probability of observing a fragmenting disc with current technology is very low -- although it will improve with Herschel and ALMA surveys (Stamatellos et al. 2011).

In the meantime, it is appropriate to explore the possibility of forming brown-dwarf binaries other than by disc fragmentation, and then capturing them into wide orbits around Sun-like stars. This paper reports numerical experiments which suggest that this is unlikely. In Section \ref{SEC:ICS} we define the initial conditions used for the numerical experiments. In Section \ref{SEC:NUM} we describe the numerical method. In Section \ref{SEC:DEF} we define the categories used in interpreting the results. Section \ref{SEC:RES} presents the results, and Section \ref{SEC:DIS} presents discussion of the results and summarises our conclusions.

\section{Initial conditions}\label{SEC:ICS}

We follow the interaction of three point-mass stars, labelled 1, 2, 3. The first star is a Sun-like star with mass $M_{_1}=1\,{\rm M}_\odot$, and the other two are brown dwarfs with masses $M_{_2}=M_{_3}=0.05\,{\rm M}_\odot$.

At the start of a simulation, the two brown dwarfs (Stars 2 and 3) are in a close binary system, on a circular orbit with separation $S_{_{23}}=5\,{\rm AU}$, hence period
\begin{eqnarray}
P_{_{23}}&=&2\pi\,\left(\frac{S_{_{23}}^3}{2GM_{_2}}\right)^{1/2}\;\,\simeq\;\, 35\,{\rm years}\,,
\end{eqnarray}
orbital velocity
\begin{eqnarray}
v_{_{23}}&=&\left(\frac{GM_{_2}}{2S_{_{23}}}\right)^{1/2}\;\,\simeq\;\, 2.1\,{\rm km}\,{\rm s}^{-1}\,,
\end{eqnarray}
and net energy
\begin{eqnarray}
E_{_{23}}&=&-\,\frac{GM_{_2}M_{_3}}{2S_{_{\rm 23}}}\;\,\simeq\;\,-\,4.6\times 10^{42}\,{\rm erg}\,.
\end{eqnarray}
These are representative parameters for a browm-dwarf binary in the field. The centre of mass of this binary is placed at the origin of coordinates. The orientation and phase of the binary orbit are random. 

At the start of a simulation, the Sun-like star (Star 1) has position $(x,y,z)=(b,-3000,0)\,{\rm AU}$ with $b$ chosen randomly from
\begin{eqnarray}
p_b\,db&=&\frac{2\,b\,db}{\left(3000\,{\rm AU}\right)^2}\,,\hspace{1cm}0\,{\rm AU}<b<3000\,{\rm AU}\,;
\end{eqnarray}
and velocity $(v_x,v_y,v_z)=(0,|v|,0)\,{\rm km}\,{\rm s}^{-1}$ with $v$ chosen randomly from
\begin{eqnarray}\label{EQN:VD}
p_v\,dv&=&\frac{1}{(2\pi)^{1/2}\,\sigma}\,\exp\left\{\frac{-\,v^2}{2\,\sigma^2}\right\}\,dv\,,
\end{eqnarray}
with $\sigma=1,\,2,\,3,\,4\,{\rm km}\,{\rm s}^{-1}$. Hence the additional system energy introduced by the Sun-like star is
\begin{eqnarray}
E_{_1}&\simeq&-\;\frac{GM_{_1}\left(M_{_2}+M_{_3}\right)}{\left((3000\,{\rm AU})^2+b^2\right)^{1/2}}\,
+\,\frac{M_{_1}\left(M_{_2}+M_{_3}\right)v^2}{2\left(M_{_1}+M_{_2}+M_{_3}\right)}\nonumber\\
&\simeq&-\,0.3\times 10^{42}\,{\rm erg}\,\left(1+\left(\frac{b}{3000\,{\rm AU}}\right)^{\!2}\right)^{\!-1/2}\nonumber\\
&&\hspace{1.4cm}+\;0.9\times 10^{42}\,{\rm erg}\left(\frac{v}{{\rm km}\,{\rm s}^{-1}}\right)^2\,.
\end{eqnarray}
The velocity dispersions are representative of those in young star clusters, and we are assuming that the velocities are random and isotropic, with a Gaussian distribution. The distribution of impact parameters takes account of the fact that the integrated cross-section at impact parameter $b$ is $\pi b^2$; the range of impact parameters is that for which significant interactions between the Sun-like star and the brown dwarf binary can be expected. The starting positions are chosen so that gravitational focussing is captured. We note that, although the approach of the Sun-like star is in the $z=0$ plane, these are not coplanar interactions, since the orientation of the orbit of the brown dwarf binary is random.

\begin{table*}
\small
\begin{center}
\caption{Abbreviations for different outcomes, and their definitions.\label{TAB:DEFS}}
\begin{tabular}{l|l}\tableline\tableline
{\sc Abbreviation}$\;\;\;$ & {\sc Definition} \\\tableline
HT1 & {\bf H}ierarchical {\bf T}riple of Type {\bf 1} \\
    & (close system contains both brown dwarfs); \\
HT2 & {\bf H}ierarchical {\bf T}riple of Type {\bf 2} \\
    & (close system contains the Sun-like star);\\
NHT & {\bf N}on-{\bf H}ierarchical {\bf T}riple; \\
X   & e{\bf X}change (one brown dwarf captured \\
    &  by Sun-like star, other ejected);\\
S   & {\bf S}urvival (brown-dwarf binary survives, \\
    & but not captured by Sun-like star); \\
D   & {\bf D}isintegration (all stars become single, \\
    & only possible with extreme $v$). \\\tableline
\end{tabular}
\end{center}
\end{table*}

In our numerical experiments, we only vary five parameters: two angles, $(\theta,\phi)$, giving the orientation of the orbit of the brown-dwarf binary; a third angle, $\psi$, giving the phase of one of the brown dwarfs in its orbit; and the impact parameter, $b$, and velocity, $v$, of the incoming Sun-like star. We do not consider different masses for the stars $(M_1,M_2,M_3)$, nor do we consider close brown-dwarfs binaries with different initial orbital parameters (separations, $S_{23}$, and eccentricities, $e_{23}$). The reason for limiting our numerical experiments to the first five parameters is that the outcome of an interaction depends in a chaotic manner on the orientation of the brown-dwarf orbit, the phase of the brown dwarfs in this orbit at the point of closest approach of the incoming Sun-like star, and the velocity and distance of the incoming Sun-like star at the point of closest approach. If one of the last five parameters were changed, the areas of this phase-space corresponding to different generic outcomes (see Section \ref{SEC:RES} and Table \ref{TAB:DEFS}) would shift, but the overall topology of the phase-space and the relative frequencies of different outcomes would change very little. In other words, if the Sun-like star considered here almost never captures the close brown-dwarf binary considered here, it is extremely unlikely that Sun-like stars of different (but comparable) mass would capture brown-dwarf binaries with different (but comparable) properties.

We reiterate that the collision parameters are chosen to reflect the typical velocity dispersions in star clusters ($1\,{\rm km}\,{\rm s}^{-1}\la\sigma\la 4\,{\rm km}\,{\rm s}^{-1}$) and the range of impact parameters ($0\la b\la 3000\,{\rm AU}$) for which a close interaction is {\it possible, but not guaranteed}, at these velocities. In effect, this means exploring impact parameters and starting distances at which the gravitational potential energy of the brown-dwarf binary relative to the Sun-like star is less than their relative kinetic energy. Under this circumstance, there are many collisions that are simply flybys, and leave the brown-dwarf binary intact (outcome S in Table \ref{TAB:DEFS}). Obviously our main concern is with the non-flybys.

For each value of $\sigma$ we perform a Monte-Carlo ensemble of $10^6$ realisations, each of which is followed for $150\,{\rm kyr}$. A single realisation is defined by the five random numbers used to generate $\theta$, $\phi$, $\psi$, $v$ and $b$.

\begin{figure*}
\epsscale{.80}
(a)\includegraphics[angle=-90, width=0.469\textwidth]{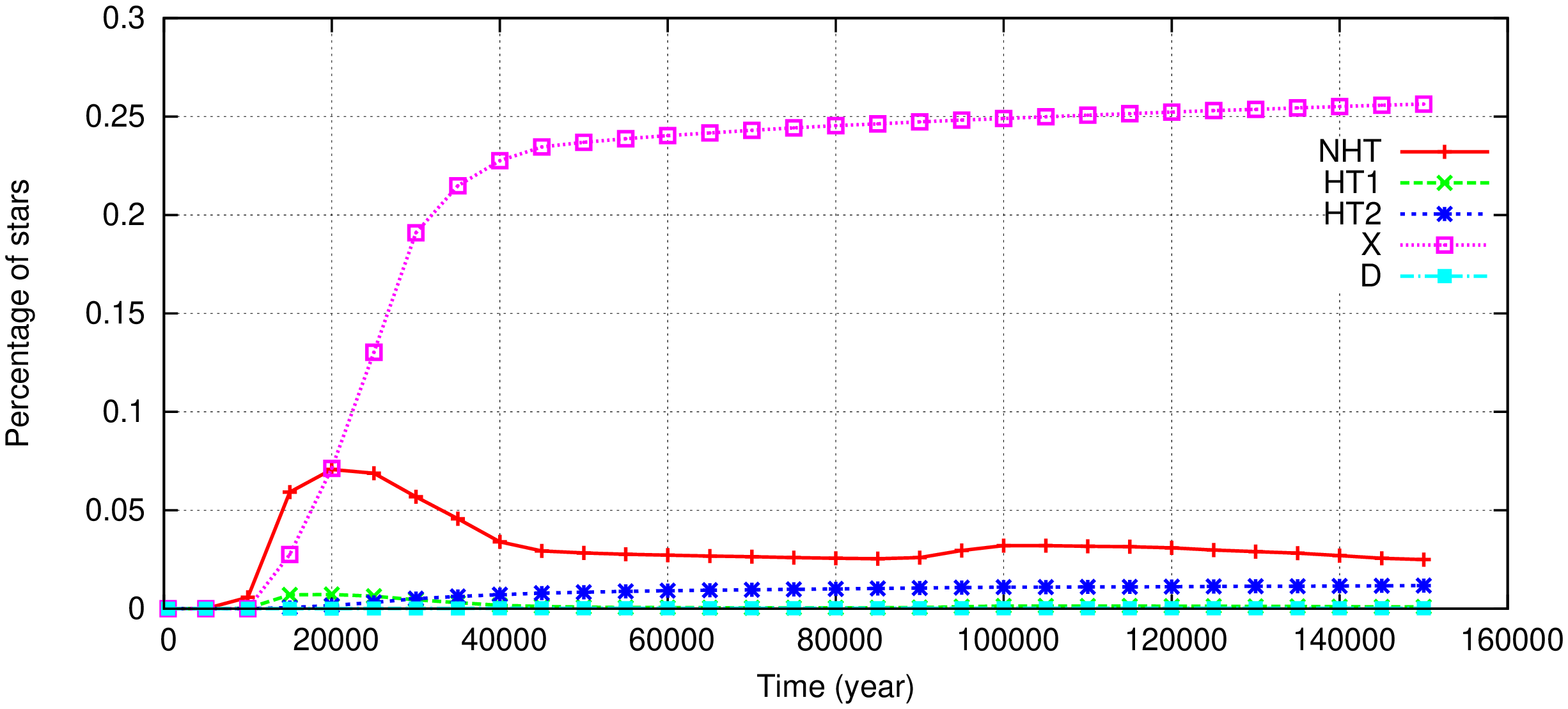} 
\includegraphics[   angle=-90, width=0.469\textwidth]{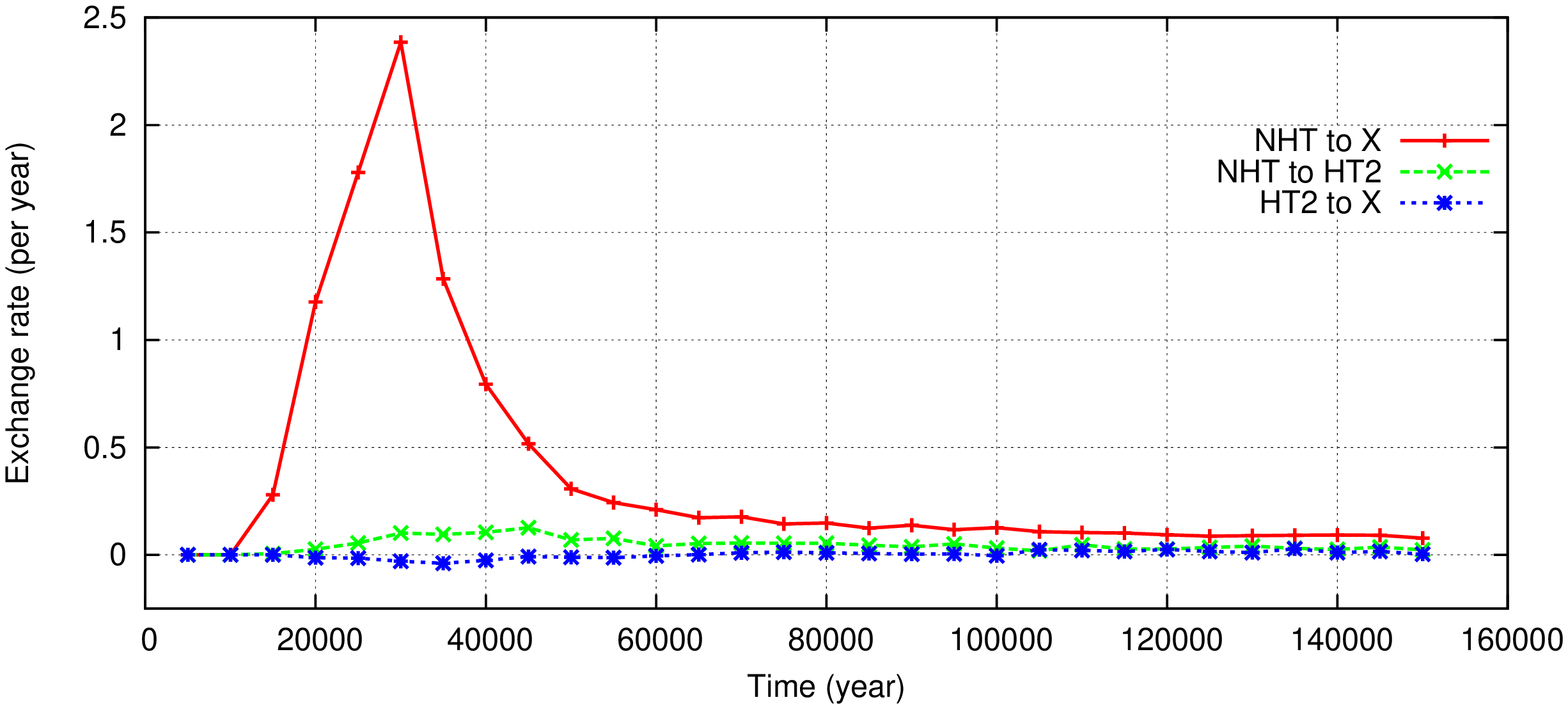}(b) \\\vspace{0.3cm}
(c)\includegraphics[angle=-90, width=0.469\textwidth]{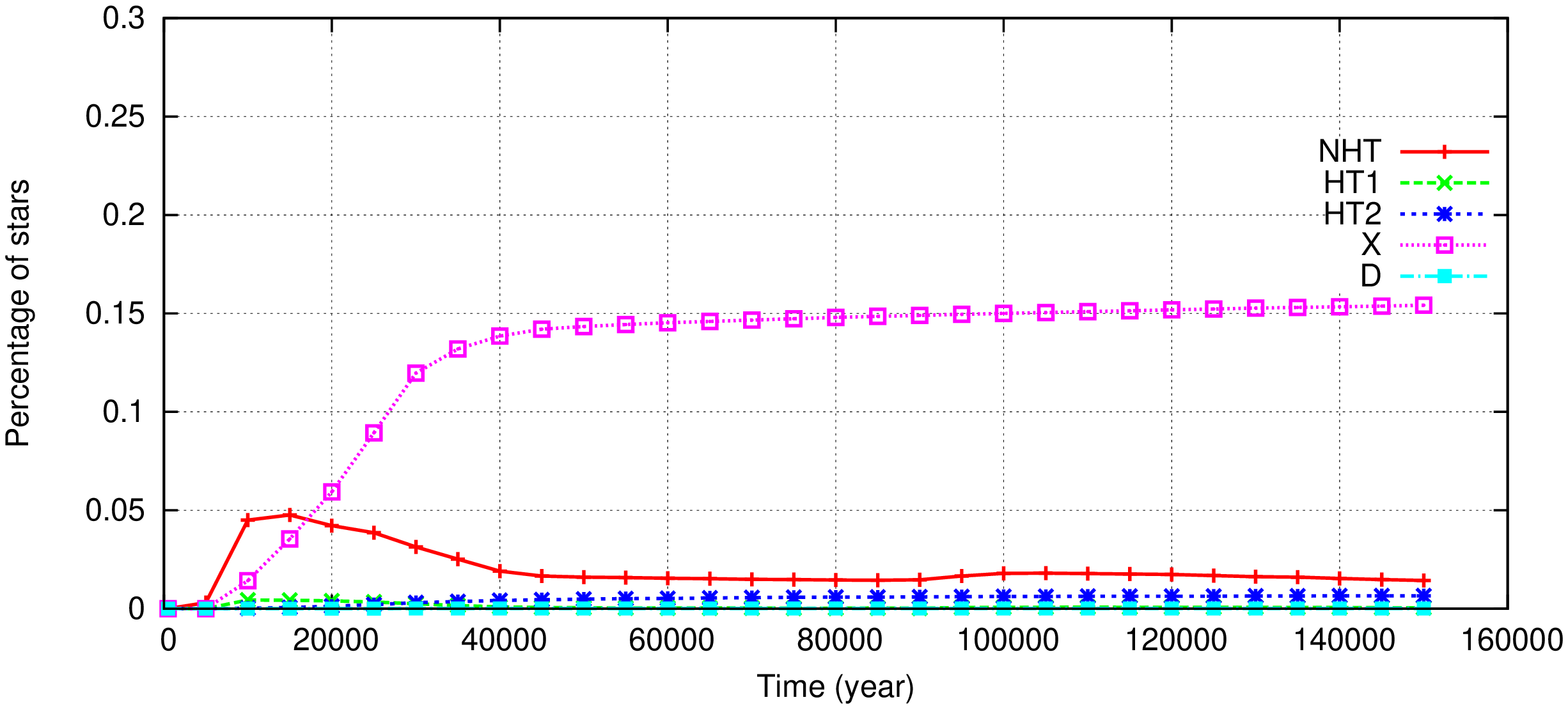} 
\includegraphics[   angle=-90, width=0.469\textwidth]{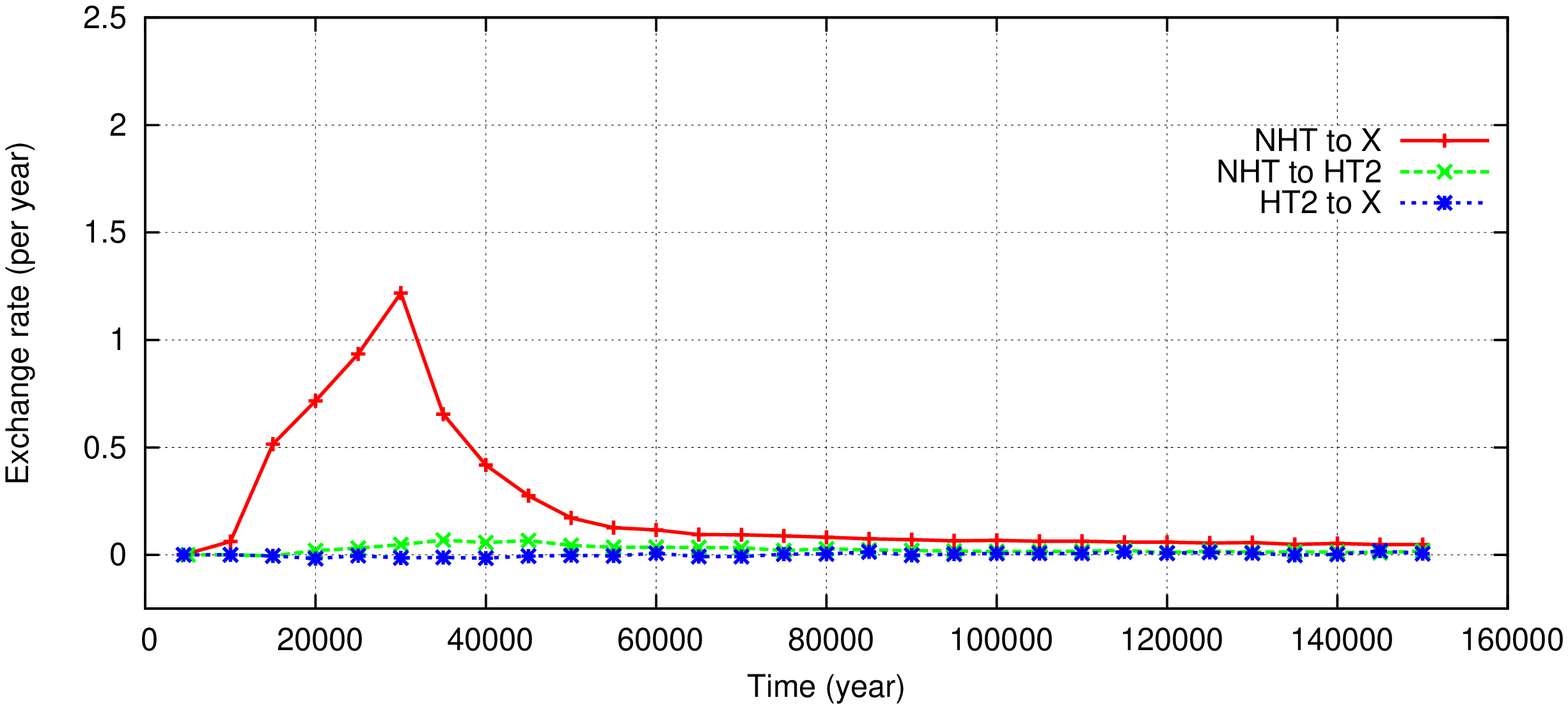}(d) \\\vspace{0.3cm}
(e)\includegraphics[angle=-90, width=0.469\textwidth]{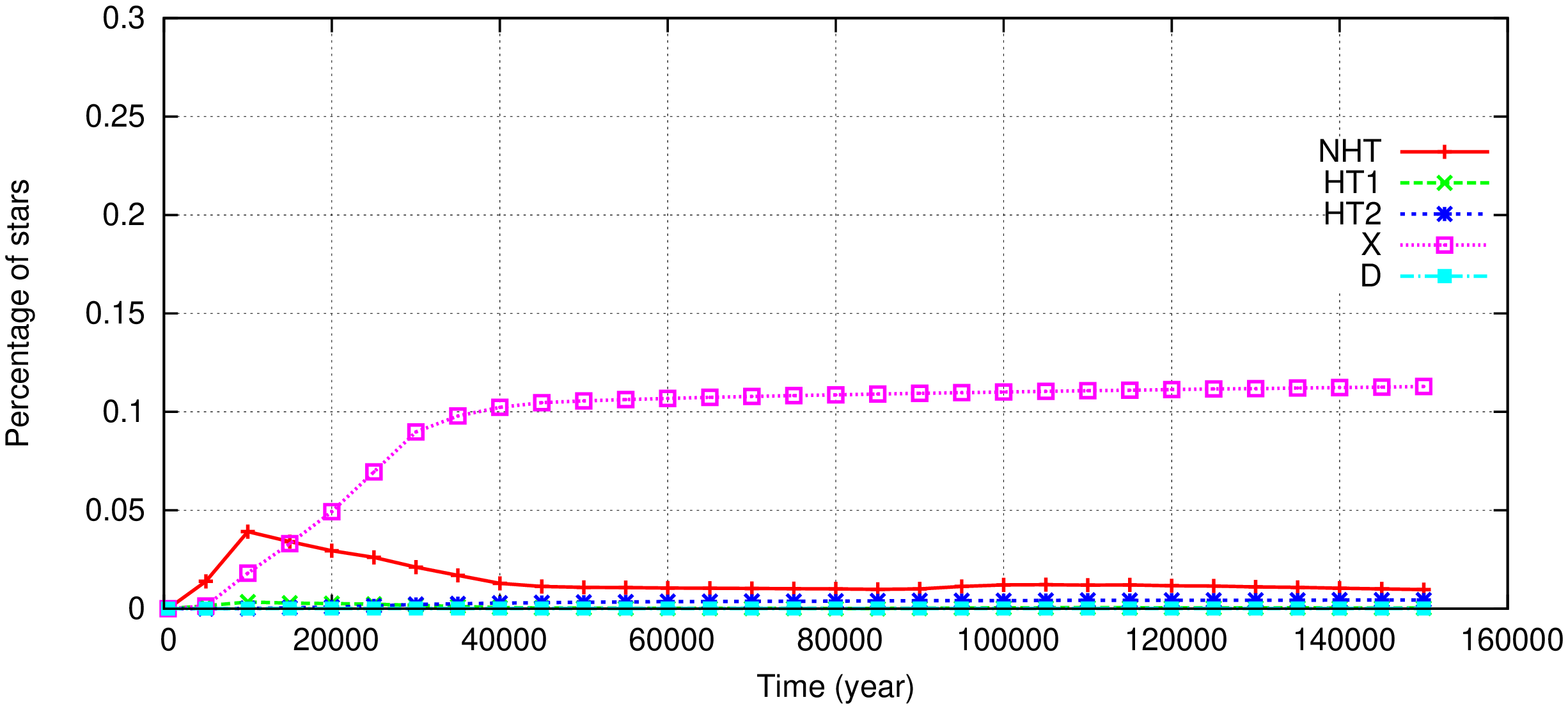} 
\includegraphics[   angle=-90, width=0.469\textwidth]{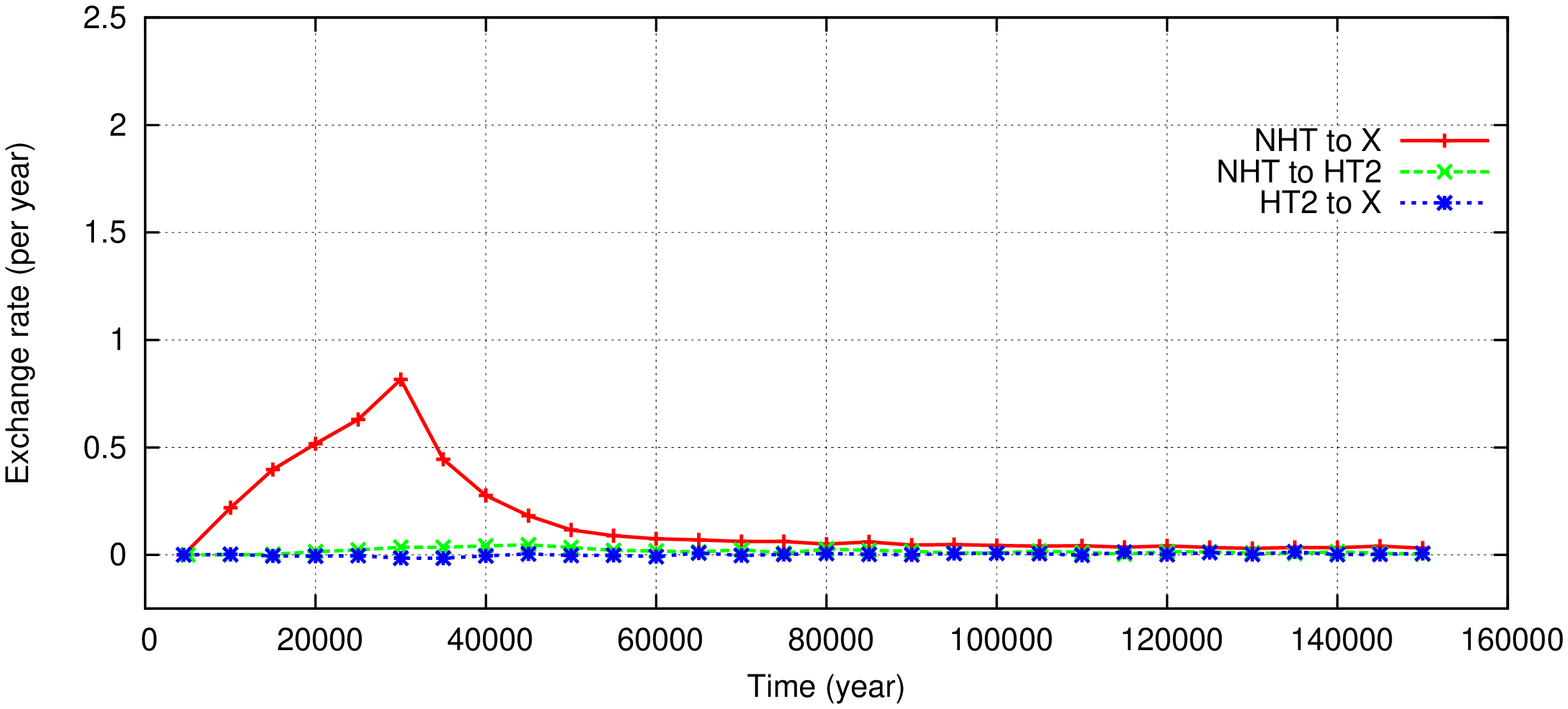}(f) \\\vspace{0.3cm}
(g)\includegraphics[angle=-90, width=0.469\textwidth]{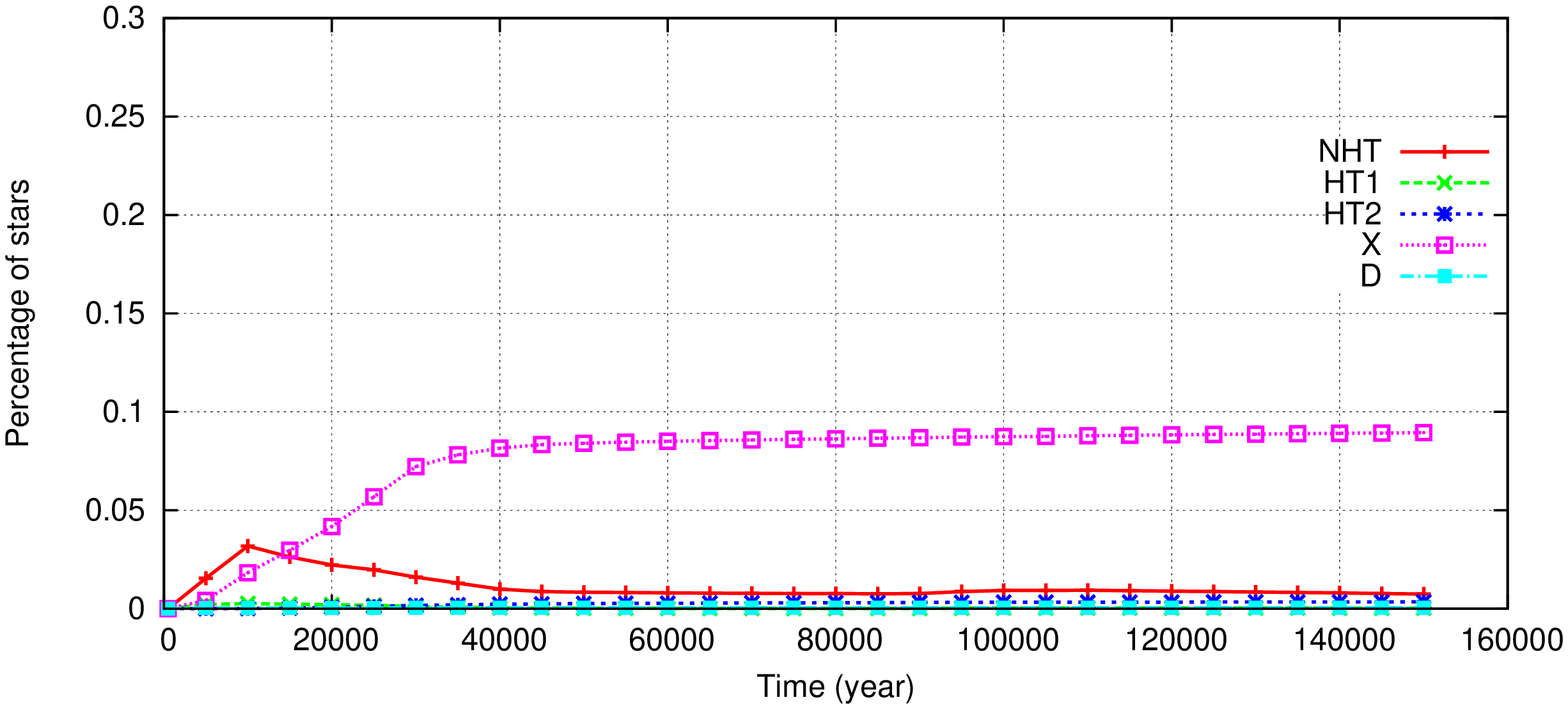} 
\includegraphics[   angle=-90, width=0.469\textwidth]{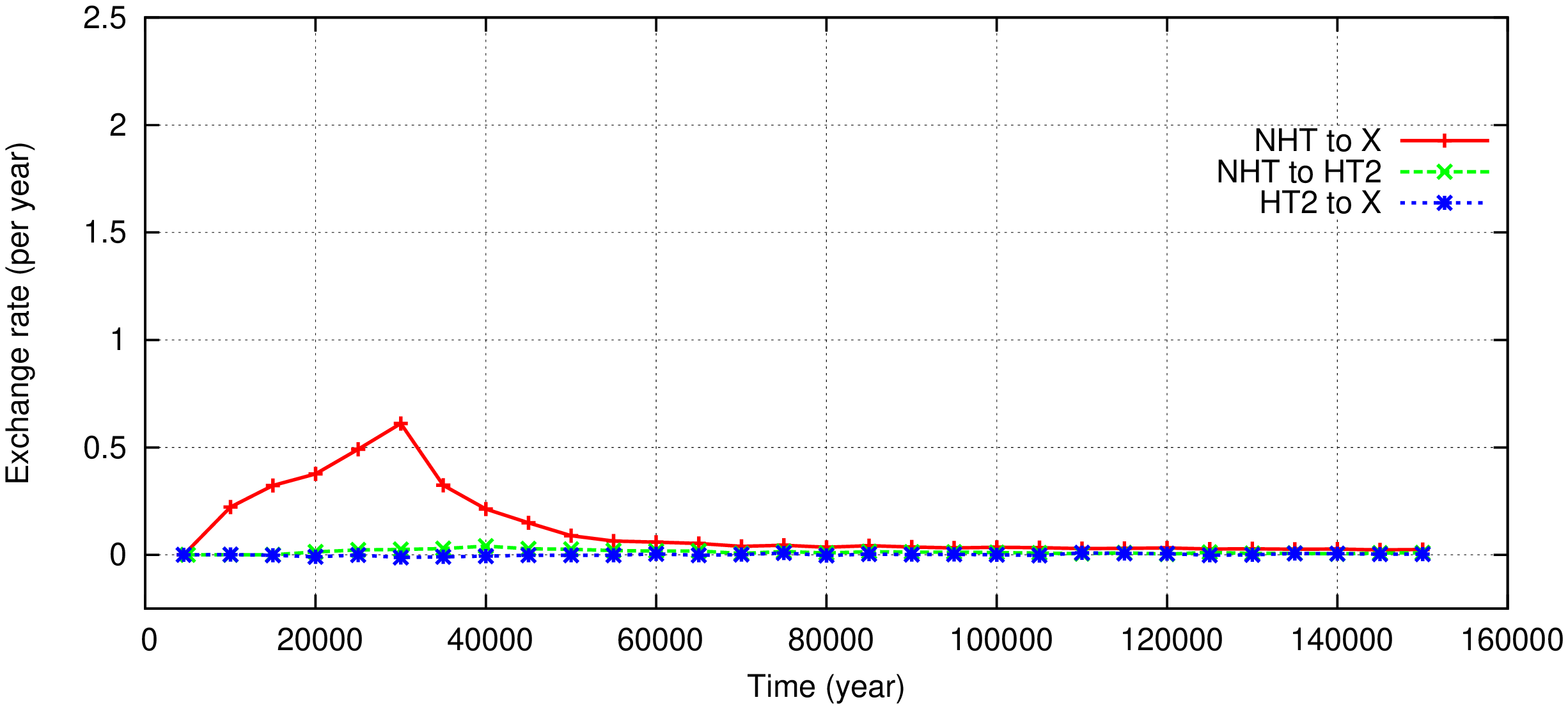}(h) \\\vspace{0.3cm}
\caption{The left hand figures give the numbers of systems in the different designations (X, NHT, HT2, HT1 and D), as a function of time; we omit designation S, since, although it is the most common outcome, it corresponds to cases where there is no significant interaction between the Sun-like star and the brown-dwarf binary. These plots show that at the end the dominant designation after S is X (one brown dwarf captured by the Sun-like star, one ejected); followed by NHT (non-hierarchical triple); followed by HT2 (hierarchical triple with one brown dwarf in a close orbit round the Sun-like star, and one in a wide orbit). There are very few HT1 systems (hierarchical triples with the brown dwarfs in a close system on a wide orbit round the Sun-like star) or D systems (total disruption). The right hand figures give the rates at which systems are exchanged between designations. This shows that the main channels are converting NHT systems into HT2 or X, and HT2 into X. In other words, in almost all cases where there is a significant interaction, one of the brown dwarfs ends up either ejected directly , or scattered into a wide orbit and then ejected, or scattered into a wide orbit. The first row presents results for $\sigma_v=1\,{\rm km}\,{\rm s}^{-1}$, the second row for $\sigma_v=2\,{\rm km}\,{\rm s}^{-1}$, the third row for $\sigma_v=3\,{\rm km}\,{\rm s}^{-1}$, and the fourth row for $\sigma_v=4\,{\rm km}\,{\rm s}^{-1}$.}
\label{FIG:MONTAGE}
\end{figure*}

\section{Numerical method}\label{SEC:NUM}

Each realisation is evolved using a fourth-order Hermite scheme with an adaptive global timestep given by
\begin{eqnarray}
\Delta t&=&\gamma\;\mbox{\sc min}\left\{\frac{|\Delta{\bf r}_{_{ij}}|}{|\Delta{\bf v}_{_{ij}}|}\,,\;\left(\frac{|\Delta{\bf r}_{_{ij}}|}{|\Delta{\bf a}_{_{ij}}|}\right)^{1/2}\right\}\,;
\end{eqnarray}
here $\Delta{\bf r}_{_{ij}}={\bf r}_{_i}-{\bf r}_{_j}$, $\Delta{\bf v}_{_{ij}}={\bf v}_{_i}-{\bf v}_{_j}$, and $\Delta{\bf a}_{_{ij}}={\bf a}_{_i}-{\bf a}_{_j}$, and the minimisation is taken over all pairs of stars, $\,ij=12,\;23,\;{\rm and}\;31$.

We use $\,\gamma=0.01\,$, and this ensures that the fractional error in the total energy, $E_{_{\rm TOT}}$, linear momentum, ${\bf P}_{_{\rm TOT}}$, and angular momentum, ${\bf H}_{_{\rm TOT}}$, satisfy
\begin{eqnarray}
\log_{_{10}}\left\{\frac{\left|\Delta E_{_{\rm TOT}}\right|}{\left|E_{_{\rm TOT}}\right|}\right\}&\simeq&-5.5\pm 1.0\,,\\
\log_{_{10}}\left\{\frac{\left|\Delta{\bf P}_{_{\rm TOT}}\right|}{\left|{\bf P}_{_{\rm TOT}}\right|}\right\}&\simeq&-13.3\pm 1.2\,,\\
\log_{_{10}}\left\{\frac{\left|\Delta{\bf H}_{_{\rm TOT}}\right|}{\left|{\bf H}_{_{\rm TOT}}\right|}\right\}&\simeq&-8.6\pm 1.5\,.
\end{eqnarray}
These uncertainties can be reduced further, but this would not significantly alter the statistics of the ensemble.

\section{Definitions}\label{SEC:DEF}

At any time, and for each pair, $ij$, we can compute the total mass, $M_{ij}$, centre-of-mass position, ${\bf r}_{ij}$, centre-of-mass velocity, ${\bf v}_{ij}$ and mutual energy, $E_{ij}$, according to
\begin{eqnarray}
M_{ij}&=&M_i\;+\;M_j\,,\\
{\bf r}_{ij}&=&\frac{M_i{\bf r}_i\,+\,M_j{\bf r}_j}{M_{ij}}\,,\\
{\bf v}_{ij}&=&\frac{M_i{\bf v}_i\,+\,M_j{\bf v}_j}{M_{ij}}\,,\\
E_{_{ij}}&=&-\;\frac{GM_iM_j}{\left|{\bf r}_i-{\bf r}_j\right|}\;+\;\frac{M_iM_j\left|{\bf v}_i-{\bf v}_j\right|^2}{2\left(M_i+M_j\right)}\,.
\end{eqnarray}
We can also compute the energy of the third star, $k$, relative to this pair (treated, rightly or wrongly, as an unresolved pair):
\begin{eqnarray}
E_k&=&-\;\frac{GM_{ij}M_k}{\left|{\bf r}_{ij}-{\bf r}_k\right|}\;+\;\frac{M_{ij}M_k\left|{\bf v}_{ij}-{\bf v}_k\right|^2}{2\left(M_{ij}+M_k\right)}\,.
\end{eqnarray}
If
\begin{eqnarray}\label{EQN:ESC}
E_k&>&0\,,
\end{eqnarray}
we label $k$ as an escaper (but see below). If $k=1$, we designate the system S, meaning that the Sun-like star has remained single, and the brown-dwarf binary has survived intact (albeit it with an adjusted orbit). If $k\neq 1$, we designate the system X, meaning that there has been an exchange, one of the brown dwarfs has been ejected and the other has been captured by the Sun-like star. If none of the $E_k$ satisfies Inequality (\ref{EQN:ESC}), but one of the pairs satisfies
\begin{eqnarray}\label{EQN:HT}
E_{ij}&<&10\,{\rm MIN}\!\left(E_{ik},E_{jk}\right)\,,
\end{eqnarray}
we label $ij$ as a close pair (but again see below), and designate the system hierarchical. The factor $10$ is somewhat arbitrary, but the results do not depend strongly on the number chosen here; the larger the number, the more stable are the hierarchical systems. We distinguish two types of hierarchical system: In HT1 systems, the close pair is $ij=23$, and therefore the Sun-like star is presumed to have captured the brown-dwarf binary into a (relatively) wide orbit; in HT2 systems, the close pair is $ij=12$ or $ij=13$, and therefore the Sun-like star has captured one of the brown dwarfs into a close orbit, and scattered the other one into a wide orbit. If neither Inequality (\ref{EQN:ESC}) nor Inequality (\ref{EQN:HT}) are satisfied, the system is designated NHT for non-hierarchical triple. We summarise these categories in Table \ref{TAB:DEFS}.

These labels and designations are not rigorous, in the following sense. There are transient configurations in which star $k$ satisfies Inequality (\ref{EQN:ESC}) but is not really an escaper, or pair $ij$ satisfies Inequality (\ref{EQN:HT}) but the system in not really hierarchical. For example, one of the stars in a close binary system may have such a large orbital velocity, and hence such a large velocity relative to the third star, that it appears to be an escaper, even though the binary as a whole is bound to the third star, {\it and vice versa}. However, these circumstances are both short-lived and increasingly rare, so that once the system has been evolved for $150\,{\rm kyr}$ (i.e. more than 4000 times the initial binary period), they have a negligible impact on the statistics. 

\begin{table}
\begin{center}
\caption{Relative frequencies of different outcomes.\label{TAB:END}}
\begin{tabular}{c|cccc}\tableline\tableline
$\sigma_v/({\rm km}/{\rm s})$ & 1 & 2 & 3 & 4 \\\tableline
$f_{_{\rm S}}$ & 0.7069 & 0.8243 & 0.8726 & 0.8994 \\
$f_{_{\rm X}}$ & 0.2564 & 0.1541 & 0.1129 & 0.0895 \\
$f_{_{\rm NHT}}$ & 0.0249 & 0.0143 & 0.0097 & 0.0074 \\
$f_{_{\rm HT2}}$ & 0.0117 & 0.0066 & 0.0044 & 0.0035 \\
$f_{_{\rm HT1}}$ & 0.0009 & 0.0005 & 0.0003 & 0.0002 \\
$f_{_{\rm D}}$ & 0.0002 & 0.0002 & 0.0001 & 0.0000 \\\tableline
\end{tabular}
\end{center}
\end{table}

\section{Results}\label{SEC:RES}

Fig. \ref{FIG:MONTAGE} presents the results obtained with $\sigma=1\,{\rm km}\,{\rm s}^{-1}$ (first row),  $2\,{\rm km}\,{\rm s}^{-1}$ (second row), $3\,{\rm km}\,{\rm s}^{-1}$ (third row) and $4\,{\rm km}\,{\rm s}^{-1}$ (fourth row). On each row, the left panel shows the number of systems in each category apart from S (see Table \ref{TAB:DEFS}), as a function of time, and the right panel shows the rates at which systems are transfered from one category to another (only the dominant channels, i.e. NHT$\rightarrow$X; NHT$\rightarrow$HT2; HT2$\rightarrow$X), again as a function of time.

Table \ref{TAB:END} summarises the final outcome in terms of the fraction of experiments that produce systems in the different categories after 150,000 years. The dominant category is S (survival, i.e. the brown-dwarf binary survives and is not captured by the Sun-like star), followed by X (exchange, i.e. one brown dwarf is captured by the Sun-like star and the other is ejected), followed by NHT (non-hierarchical triple), followed by HT2 (hierarchical triple of type 2, i.e. one brown dwarf captured into a close orbit around the Sun-like star, and the second brown dwarf captured into a wide orbit around the Sun-like star), followed by HT1 (hierarchical triple of type 1, i.e. the close brown-dwarf binary survives as such but is captured into a wide orbit around the Sun-like star), followed by D (total disruption, i.e. three single stars). Thus the majority of systems that experience a significant interaction undergo exchange and end up as X systems, i.e. one of the brown dwarfs is captured by the Sun-like star, and the other is ejected. There is an extremely low probability of producing HT1 systems, i.e. systems like the observed ones with a close brown-dwarf binary on a wide orbit around a Sun-like star. Furthermore, for each such system produced, there are many more HT2 and X systems produced, i.e. hierarchical systems with one brown dwarf in a close orbit around the Sun-like star and the other on a much wider orbit, and exchange systems where one of the brown dwarfs has been ejected.

As the velocity dispersion, $\sigma_v$, is  increased, the approach velocity of the Sun-like star increases, and the influence of gravitational focussing diminishes, so there are fewer significant interactions. Consequently $df_{_{\rm S}}/d\sigma_v>0$, but for all other outcomes (O = X, NHT, HT1, HT2, D) $df_{_{\rm O}}/d\sigma_v<0$.

\section{Discussion and conclusions}\label{SEC:DIS}

The numerical experiments described above suggest that there is an extremely low probability that a Sun-like star captures a pre-existing close brown-dwarf binary into a wide orbit, if the interaction between them is purely gtavitational. If there is such an interaction, the end result is almost always that one brown dwarf is either ejected, or scattered into a wide orbit, whilst the other becomes a close companion to the Sun-like star.

One might argue that, if the Sun-like star and/or the brown-dwarf binary were attended by circumstellar and/or circumbinary discs, interactions between them would be more dissipative and it might then be possible for the Sun-like star to capture the brown-dwarf binary into a wide orbit. However, disc dissipation would require a rather close interaction, and this would be likely to leave the brown-dwarf binary on an eccentric orbit with a close periastron to the Sun-like star. With the disc now dissipated, the simulations presented here indicate that it is then virtually inevitable that one of the brown dwarfs will end up either being ejected, or being scattered into a wide orbit, whilst the other becomes a close companion to the Sun-like star. This would be in direct conflict with the observed lack of brown dwarfs orbiting close to Sun-like stars (The Brown Dwarf Desert). 

Furthermore, if close brown-dwarf binaries retain dissipative discs long enough for them to regulate close interactions with Sun-like stars, it seems likely that single brown dwarfs do too. However, this will make it easy for single brown dwarfs to be captured into close orbits around Sun-like stars, and again this will be in conflict with The Brown Dwarf Desert.

Another possibility is that a brown dwarf binary might be delivered into a wide orbit around a Sun-like star by an exchange reaction, i.e. a Sun-like star which already had some other sort of companion might swap this companion for a brown-dwarf binary in a four-body interaction. We cannot comment quantitatively on this possibility on the basis of the results presented here (which only involve three bodies), but it seems rather contrived and unlikely. First, such exchanges normally work in the direction of swapping a companion having a low mass for one having a higher mass; therefore this route requires some other mechanism to form the initial binary system with a Sun-like primary and a very low-mass secondary, as a prelude to swapping the very low-mass secondary for a brown dwarf binary. Second, in an environment where exchanges occur (i) such exchanges will also act to remove a brown dwarf binary bound to a Sun-like star, in favour of an even more massive companion, and (ii) non-exchange interactions will scatter brown dwarf binaries that remain bound to Sun-like primaries into a range of orbits, thereby populating The Brown Dwarf Desert.

We conclude that close brown-dwarf binaries in wide orbits around Sun-like stars are likely to have been formed there, by disc fragmentation, and that if these brown dwarfs formed by disc fragmentation, then other brown dwarfs (those not in close binary systems and/or no longer in orbit around Sun-like stars) may also have formed by disc fragmentation. We note that simulations of disc fragmentation reproduce all the discriminating statistical properties of brown dwarfs (Stamatellos \& Whitworth 2009), and we advocate that alternative theories of brown dwarf formation be simulated with sufficient resolution to enable the same comparisons to be made.

\acknowledgments
MK thanks TUBITAK (Grant Num.:B.02.1.TBT.\-0.06.\-01.219.01-12-14) for financial support, which enabled this work to be undertaken. DS and APW acknowledge the support of STFC grant ST/H001530/1.

\label{lastpage}

\begin{thebibliography}{}%
\bibitem[]{}Ahmic, M., Jayawardhana, R., Brandeker, A., Scholz, A., van Kerkwijk, M.H., Delgado-Donate, E., Froebrich, D., 2007, ApJ, 671, 2074
\bibitem[]{}Bate, M.R., 1998, ApJ, 508, L95
\bibitem[]{}Bate, M.R., 2009, MNRAS, 392, 1363
\bibitem[]{}Bate, M.R., Bonnell, I.A., Bromm, V., 2003, MNRAS, 339, 577
\bibitem[]{}Bonnell, I.A., Clark, P.C., Bate, M.R., 2008, MNRAS, 389, 1556
\bibitem[]{}Bouy, H., Brandner, W., Mart{\'i}n, E.L., Delfosse, X., Akkard, F., Basri, G., 2003, AJ, 126, 1526
\bibitem[]{}Burgasser, A.J., Kirkpatrick, J.D., Lowrance, P.J., 2005, AJ, 129, 2849
\bibitem[]{}Burgasser, A.J., Kirkpatrick, J.D., Reid, I.N., Brown, M.E., Miskey, C.L., Gizis, J.E., 2003, ApJ, 586, 512
\bibitem[]{}Close. L.M., Siegler, N., Freed, M., Biller, B., 2003, ApJ, 587, 407
\bibitem[]{}Comer{\'o}n, F., Testi, L., Natta, A., 2010, A\&A, 522, 47
\bibitem[]{}Faherty, J.K., Burgasser, A.J., Bochanski, J.J., Looper, D.L., West, A.A., van der Bliek, N.S., 2011, AJ, 141, 71
\bibitem[]{}Faherty, J.K., Burgasser, A.J., West, A.A., Bochanski, J.J., Cruz, K.L., Shara, M.M., Walter, F.M., 2010, AJ, 139, 176
\bibitem[]{}Gammie, C.F., 2001, ApJ, 553, 174
\bibitem[]{}Gelino, C.R., Burgasser, A.J., 2010, AJ, 140, 110
\bibitem[]{}Girichidis, P., Federrath, C., Banerjee, R., Klessen, R.S., 2010, MNRAS in press
\bibitem[]{}Goodwin, S.P., Whitworth, A.P., Ward-Thompson, D., 2004a, A\&A, 414, 633
\bibitem[]{}Goodwin, S.P., Whitworth, A.P., Ward-Thompson, D., 2004b, A\&A, 423, 169
\bibitem[]{}Hennebelle, P., Chabrier, G., 2008, ApJ, 684, 395
\bibitem[]{}Hennebelle, P., Chabrier, G., 2009, ApJ, 702, 1428
\bibitem[]{}Kratter, K.M., Matzner, C.D., Krumholz, M.R., Klein, R.I., 2010, ApJ, 708, 1585
\bibitem[]{}Kumar, S.S., 1961, AJ, 67 579
\bibitem[]{}Lodieu, N., Zapatero-Osorio, M.R., Rebolo, R., Martin, E.L., Hambly, N.C., 2009, A\&A, 505, 1115
\bibitem[]{}Luhman, K.L., 2006, ApJ, 645, 676
\bibitem[]{}Luhman, K.L., Hern{\'a}ndez, J., Downes, J.J., Hartman, L., Brice{\~n}o, C., 2008, ApJ, 688, 362
\bibitem[]{}Luhman, K.L., Joergens, V., Lada, C., Muzerolle, J., Pascucci, I., White, R., 2007, {\it Protostars and Planets V} (Eds. B. Reipurth, D. Jewitt, K. Keil; University of Arizona Press, Tucson) p.443
\bibitem[]{}McCarthy, C., Zuckerman, B., 2004, AJ, 127, 2871
\bibitem[]{}Maury, A.J., Andr{\'e}, Ph., Hennebelle, P., Motte, F., Stamatellos, D., Bate, M.R., Belloche, A., Duch{\^e}ne, G., Whitworth, A.P., 2010, A\&A, 512, 40
\bibitem[]{}Mohanty, S., Jayawardhana, R., Basri, G., 2005, ApJ, 626, 498
\bibitem[]{}Moraux, E., Bouvier, J., Stauffer, J.R., Barrado y Na\-vas\-cu{\' e}s, \- D., Cuillandre, J.-C., 2007, A\&A, 471, 499
\bibitem[]{}Nakajima, T., Oppenheimer, B.R., Kulkarni, S.R., Goli\-mows\-ki, D.A., Matthews, K., Durrance, S.T., 1995, Nature, 378, 463
\bibitem[]{}Natta, A., Testi, L., Muzerolle, J., Randich, S., Comer{\'o}n, F., Persi, P., 2004, A\&A, 424, 603
\bibitem[]{}Offner, S.S.R., Klein, R.I., McKee, C.F., Krumholz, M.R., 2009, ApJ, 703, 1310
\bibitem[]{}Offner, S.S.R., Kratter, K.M., Matzner, C.D., Krumholz, M.R., Klein, R.I., 2010, ApJ, 725, 1485
\bibitem[]{}Padoan, P., Nordlund, {\AA}., 2002, ApJ, 576, 870
\bibitem[]{}Rebolo, R., Zapatero-Osorio, M.R., Martin, E.L., 1995, Nature 377, 129
\bibitem[]{}Reid, I.N., Cruz, K.L., Burgasser, A.J., Liu, M.C., 2008, AJ, 135, 580
\bibitem[]{}Reipurth, B., Clarke, C.J., 2001, AJ, 122, 432
\bibitem[]{}Sahlmann, J., et al., 2011, A\&A, 525, 95
\bibitem[]{}Stamatellos, D., Hubber, D.A., Whitworth, A.P., 2007, MNRAS, 382, L30
\bibitem[]{}Stamatellos, D., Hubber, D.A., Whitworth, A.P., 2011, ApJ, 730, 32
\bibitem[]{}Stamatellos, D., Whitworth, A.P., 2009, MNRAS, 392, 413
\bibitem[]{}Stamatellos, D., Maury, A., Whitworth, A.P., Andr{\'e}, Ph., 2011, MNRAS, 413, 1787
\bibitem[]{}Thies, I., Kroupa, P., 2007, ApJ, 671, 767
\bibitem[]{}Thies, I., Kroupa, P., 2008, MNRAS, 390, 1200
\bibitem[]{}Todorov, K., Luhman, K.L., McLeod, K.K., 2010, ApJ, 714, L84
\bibitem[]{}Toomre, A., 1964, ApJ, 139, 1217
\bibitem[]{}Walch, S.K., Naab, T., Whitworth, A.P., Burkert, A., Gritschneder, M., 2010, MNRAS, 402, 2253
\bibitem[]{}Whelan, E.T., Ray, T.P., Randich, S., Bacciotti, F., Jayawardhana, R., Testi, L., Natta, A., Mohanty, S., 2007, ApJ, 659, L45
\bibitem[]{}Whitworth, A.P., Bate, M.R., Nordlund, {\AA}., Reipurth, B., Zinnecker, H., 2007, {\it Protostars \& Planets V} (Eds. B Reipurth, D Jewitt, K Keil; University of Arizona Press, Tucson), p.459
\bibitem[]{}Whitworth, A.P., Stamatellos, D., 2006, A\&A, 458, 817
\bibitem[]{}Whitworth, A.P., Zinnecker, H., 2004, A\&A, 427, 299
\end{thebibliography}
\end{document}